\documentclass[a4paper, preprint,12pt,authoryear]{elsarticle}

\usepackage{amssymb}

\journal{Journal of the American Ceramic Society}

\begin{document}

\begin{frontmatter}

\title{Yielding and flow of foamed metakaolin pastes}

\author[label1]{Lucie Duclou\'e\corref{cor1}}
\cortext[cor1]{corresponding author}
\ead{lucie.ducloue@manchester.ac.uk}
\author[label2]{Olivier Pitois}
\author[label2]{Laurent Tocquer}
\author[label2]{Julie Goyon}
\author[label3]{Guillaume Ovarlez}
\address[label1]{Manchester Centre for Nonlinear Dynamics and School of Physics and Astronomy, University of Manchester,Manchester M13 9PL, United Kingdom}
\address[label2]{Universit\'e Paris-Est, Laboratoire Navier, UMR 8205 CNRS - \'Ecole des Ponts ParisTech - IFSTTAR, 77420 Champs-sur-Marne, France}
\address[label3]{Universit\'e de Bordeaux, LOF, UMR 5258 CNRS - Solvay, 33608 Pessac, France}

\begin{abstract}
Metakaolin is a broadly used industrial raw material, with applications in the production of ceramics and geopolymers, and the partial replacement of Portland cement. The early stages of the manufacturing of some of these materials require the preparation and processing of a foamed metakaolin-based slurry. In this study, we propose to investigate the rheology of a foamed metakaolin-based fresh paste by performing well-controlled experiments. We work with a non-reactive metakaolin paste containing surfactant, in which we disperse bubbles of known radius at a chosen volume fraction. We perform rheometry measurements to characterize the minimum stress required for the foamed materials to flow (yield stress), and the dissipation occurring during flow. We show that the yield stress of the foamed samples is equal to the one of the metakaolin paste, and that dissipation during flow increases quadratically with the bubble volume fraction. Comparison with yielding and flow of model foamed yield stress fluids allows us to understand these results in terms of coupling between the bubbles' surface tension and the metakaolin paste's rheology.
\end{abstract}

\begin{keyword}
metakaolin \sep bubbles \sep rheology \sep yield stress \sep viscosity



\end{keyword}

\end{frontmatter}


\section{Introduction}

Metakaolin is a dehydroxylated clay~\citep{bich2009influence} with a high pozzolanic activity, making it widely used as a raw industrial material. The alkali activation of alumina- and silica-rich metakaolin leads to the formation of a three-dimensional aluminosilicate network. The resulting product, called a geopolymer~\citep{davidovits1991geopolymers} is a strong and durable cementitious material which hardens at moderate temperature, thus being relatively low energy consuming. For that reason, geopolymers are suggested as a sustainable alternative to other materials that require processing at very high temperature, among which some ceramic parts~\citep{davidovits1991geopolymers} and Portland cement, which is the most common binder for mortar and concrete. Metakaolin has also been successfully used as a partial replacement for Portland cement in regular construction materials, with benefits including an increase in the early age strength, enhanced durability and a better resistance to chemical attacks compared to ordinary Portland cement-based materials~\citep{sabir2001metakaolin}.\\

In the context of a general industrial effort to develop building materials with lower environmental impact, foamed geopolymer concrete combines the advantages of a low energy consuming geopolymer matrix with the benefits of adding gas bubbles which will make the resulting concrete lightweight and thermally insulating~\citep{ramamurthy2009classification, zhang2014geopolymer}. There are two main routes to produce foamed geopolymer concrete: gas bubbles can either be generated by an in-situ chemical reaction~\citep{prud2011situ, masi2014comparison} or can be added mechanically, either through strong agitation that entraps air in the paste~\citep{cilla2014geopolymer} or by mixing with a foam concentrate~\citep{zhang2014geopolymer}. For high temperature applications, this foamed geopolymer can then be sintered, leading to the formation of a foamed ceramic~\citep{bell2009preparation} which can be used to filter hot gases or to make thermal insulating tiles for extreme conditions~\citep{montanaro1998ceramic}.\\

Whether producing a foamed geopolymer or a foamed ceramic, the size, volume fraction and distribution of the gas bubbles in the final foamed clay material are crucial to the application the material is designed for~\citep{glad2015highly}. Therefore, the incorporation and the preservation of the gas bubbles in the mineral paste when elaborating the material is essential. From a rheological point of view, the paste is a complex fluid, which only flows if the stress applied to it is larger than a critical value: the yield stress of the paste. Above that yield stress, the paste flows with a viscosity which depends on the flow rate applied. The interplay of the complex rheology of the mineral paste with the bubbles will determine the stability of the bubbles entrapped in the paste, but also the rheology of the foamed material. As the development of materials such as geopolymer foamed concrete is currently in the relatively early stages of technological maturity, understanding this behavior is necessary to elaborate optimized processes and materials. Those stability and rheological issues should be tackled as a complement to the study of the chemical reactions and functional properties. \\

In this paper, we propose to shed light on the flow properties of just mixed foamed clay pastes by investigating the rheological behavior of a metakaolin paste containing air bubbles. We work with a non-reactive material containing surfactant to stabilize the mechanically added air bubbles, and focus on how the behavior of the paste is modified by the addition of bubbles of a known radius at a given volume fraction. This question is for instance of interest to design efficient processes to cast or pump a geopolymer foamed concrete. The first section of this paper presents in more detail our experimental system and the rheometry procedure used to characterize it. The following section presents the results we have obtained for two rheological properties relevant for bubble stability and material workability: the minimal stress required to induce flow and the dissipation during flow of the foamed metakaolin pastes. In the third and last section, those results are discussed in the light of previous experiments done on foamed samples of simple yield stress fluids.

\vspace{0.5cm}

\section{Experimental procedure}

\label{section:Mats}
We prepare model samples of foamed metakaolin paste by dispersing bubbles of well defined radius at a chosen volume fraction in a concentrated metakaolin paste. To achieve such a level of control on our systems, we prepare a very concentrated metakaolin paste containing no bubbles, which we then mix with a separately produced monodisperse aqueous foam. This procedure, which is presented below after the details of the materials used, allows us to control the bubble size and the gas volume fraction of the foamed metakaolin samples throughout the process of mixing and characterization. The volume fraction of entrapped air, $\phi$, is varied in the range 0-50\%, which is the void volume fraction commonly found in the production of foamed geopolymer concrete by mechanical agitation or mixing with a foam~\citep{zhang2014geopolymer}. Sample preparation and characterization is done at 25$^\circ$C. \\
\subsection{Metakaolin paste}
\begin{figure}[]
\centering
\includegraphics[scale=0.13]{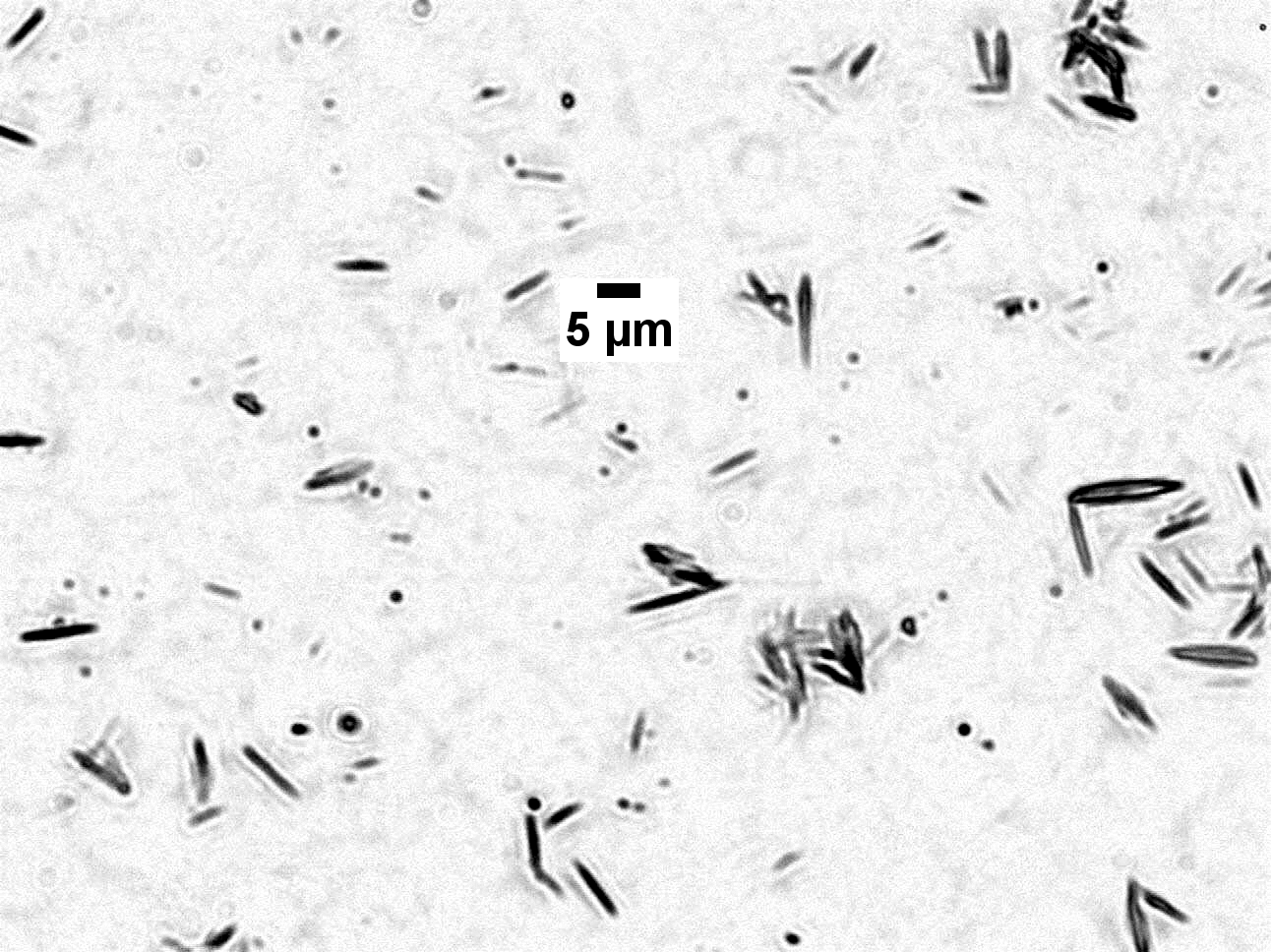}
\caption{Micrograph of a dilute suspension of M1200S in water. A drop of the suspension has been deposited on a glass slide. \label{fig:metaK}}
\end{figure}

The metakaolin paste which will be mixed with the foam ("stock paste") is a dispersion of a fine metakaolin (Argical M1200-S, AGS min\'eraux - Imerys group) powder at 46.4wt\% in an aqueous solution containing 10wt\% of a non-ionic surfactant (Tween 20$^{\textregistered}$, from Fisher Chemicals) in deionized water. The presence of this surfactant at the same concentration in the stock paste and in the aqueous foam ensures that the bubbles' surface will not be deprived of surfactants once the foam is mixed with the paste, keeping the bubbles stable to coalescence in the resulting foamed sample. A micrograph of a dilute suspension of the metakaolin powder in water in shown in figure~\ref{fig:metaK}. The grains are lamellar, and of a very small size: the median spherical equivalent diameter is 1.3 $\mu$m (data from AGS). Because of this micrometric size, there is no sedimentation nor creaming of the metakaolin suspensions in water at the concentrations we are using for the pastes. In the foamed samples, the micrometric size of the metakaolin particles will be much smaller than both the bubble radius and the typical distance between neighboring bubbles in the paste, meaning that the metakaolin paste can be described as a continuous medium embedding the bubbles. \\ 
\subsection{Monodisperse foam}
A monodisperse foam is produced separately from the paste. We use the same aqueous solution (10wt\% of Tween 20$^{\textregistered}$ in deionized water) as the one  used to prepare the stock metakaolin paste, so that the mixing of the stock paste and the foam is easy. We have used two different techniques to produce monodisperse foams with two different bubble radii. A foam containing bubbles of radius $R_b=(100 \pm 20) \mu$m has been obtained by blowing nitrogen at a few milliliters per minute through a porous glass frit. A second series of foamed suspensions containing smaller bubbles, of radius $R_b=(35 \pm 10) \mu$m has been prepared thanks to a millifluidic device (made of a T junction followed by a porous medium). In all foams, a very small amount of a gas with very low solubility in water, perfluorohexane, is added to the nitrogen to reduce coarsening~\citep{gandolfo1997interbubble}.
\subsection{Preparation of the foamed samples}
\begin{figure}[]
\centering
\includegraphics[scale=0.4]{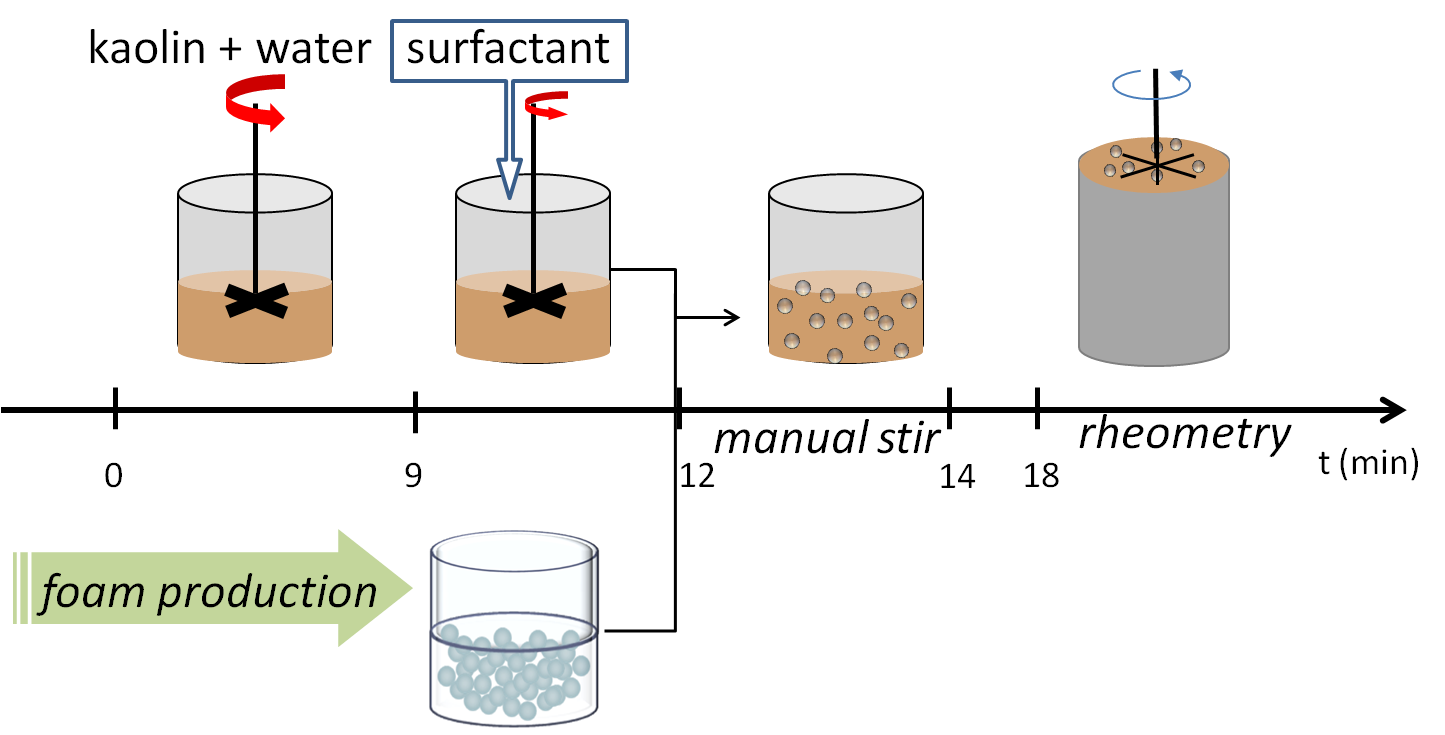}
\caption{Schematic of the timed experimental protocol (bubbles not to scale). The metakaolin powder is first dispersed in water under strong agitation, then the rotational speed is decreased and the surfactant is slowly added to the mixture. The resulting paste ("stock paste") is mixed with a separately produced monodispersed foam to get the foamed samples, which are then characterized by rheometry. \label{fig:procedure}}
\end{figure}

The protocol we use to prepare the samples and then measure their flow properties has been designed to ensure that sample preparation and characterization is reproducible. The metakaolin paste surrounding the bubbles is non-reactive, but it slowly evolves with time (like many other clays, it is thixotropic~\citep{ryan2013properties}: although no chemical bonds are created nor destroyed, the physical structure of the paste evolves slowly at rest, changing its rheological properties in response). For that reason, the preparation and measurement sequence follows a precisely timed procedure, which is detailed below and summarized schematically in figure~\ref{fig:procedure}.\\

To prepare the stock paste in a reproducible way and without entrapping unwanted air bubbles, we first disperse the powder by adding it little by little into deionized water under strong agitation. The vigorous stirring of the metakaolin suspension in water minimizes the time evolution of the resulting metakaolin paste~\citep{coussot1997rheology}. At the end of the 9 minutes necessary for the addition of metakaolin in water, we obtain a smooth and bubble-free homogeneous metakaolin paste. The speed of agitation is then decreased, and the surfactant is gradually added in the paste via a syringe. This addition of surfactant in the paste makes it prone to air entrapment, which is greatly limited by the use of a low rotation speed: measurement of the density of the prepared stock paste shows that it contains less than 1\% of air. At the end of the 3 minutes necessary to add and blend the surfactant into the paste, the paste is mixed with the separately produced foam. This mixing brings in more aqueous solution, which will lower the final solid content in the paste of the foamed samples. The mass of added foam is always 6\% of the mass of prepared stock paste, so that all the foamed metakaolin samples have the same solid content (44wt\%) in the suspending paste surrounding the bubbles ("suspending paste"). The gas volume fraction in the foamed samples is varied by changing the wetness of the aqueous foam prior to mixing. The incorporation of the bubbles in the paste is done by blending for 2 minutes with a spoon, to ensure that stirring occurs both in the vertical and the azimuthal directions. After mixing, the foamed metakaolin sample is poured in a tared rheometry cup of known volume (61.6 mL). Weighting of the full cup allows for the determination of the density of the samples, and hence the volume fraction of entrapped air. The cup is then set in the rheometer and 4 minutes after pouring the sample in the cup, the mechanical measurements start.\\

\subsection{Rheometry procedure}
\label{section:rheo}
The characterization of the flow properties of the samples is done by using a commercial controlled-stress rheometer (Bohlin C-VOR 200), which works at either set torque or set rotational velocity (thanks to a torque feedback loop). The rheometer is fitted with a six-bladed vane tool (radius $R_i=12.5$ mm) in a cup (radius $R_o=18$ mm) geometry. The gap $R_o-R_i$ is larger than several bubble diameters, which ensures that the bulk properties of the foamed samples are measured. Fine sand paper is glued inside the cup to avoid slippage of the metakaolin paste at the cup wall~\citep{coussot2005rheometry}. Because of the wide gap of the geometry we are using, the shear stress varies along the radius $r$ in the gap. We compute the shear stress and shear rate at $r=R_i$ from the torque and rotational velocity assuming that the rotating vane tool is equivalent to a cylinder of same radius $R_i$. We subtract the contribution of the end effects at the bottom of the virtual cylinder defined by the vane, and take into account the localization of shear when computing the flow curve~\citep{coussot2005rheometry}. Throughout the sequence, the free surface of the sample is visible, allowing us to check that the bubbles do not coalesce nor cream during the measurement. \\

\begin{figure}[]
\centering
\includegraphics[scale=0.235]{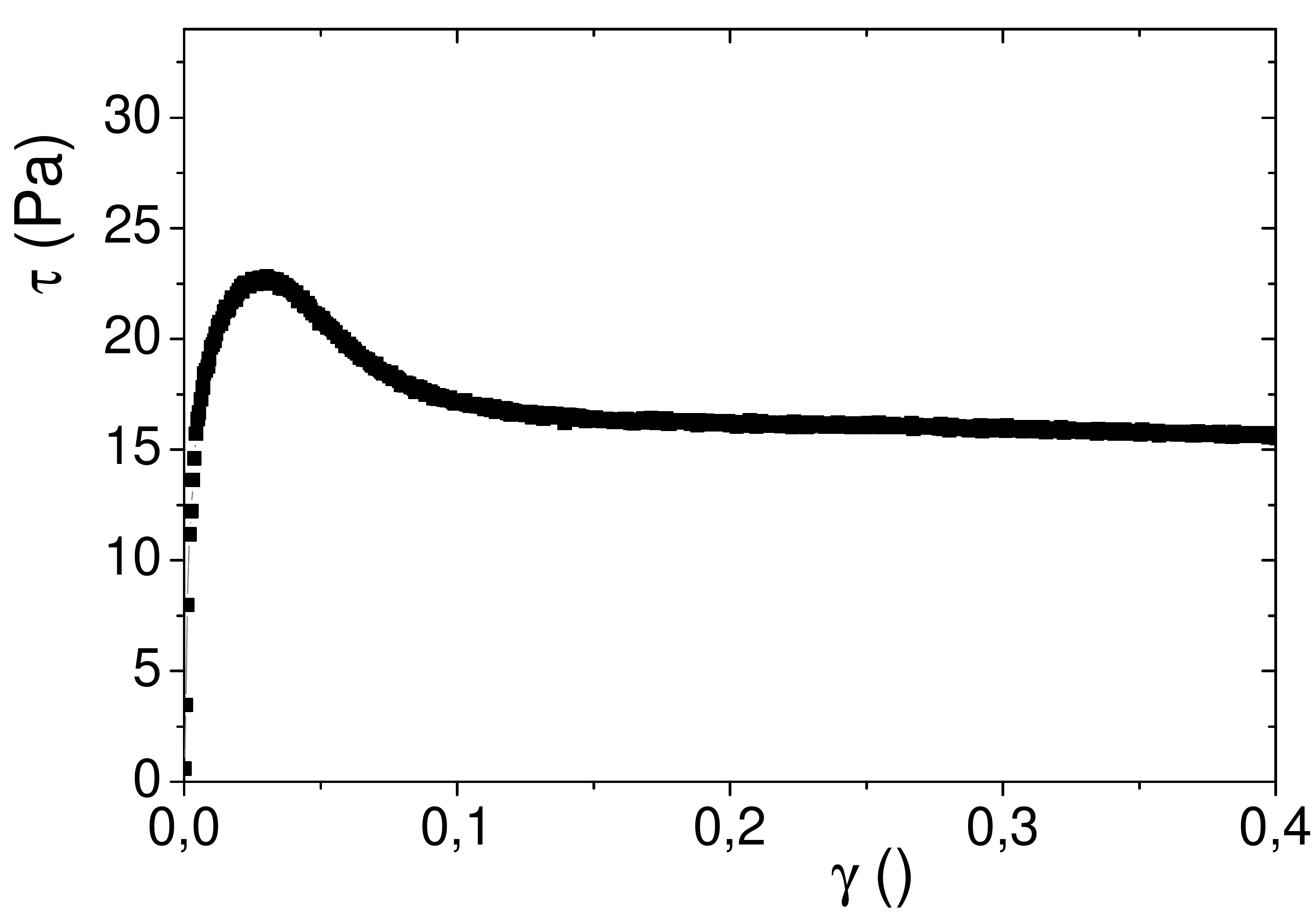}
\caption{Characterization of the suspending paste: stress as a function of strain obtained for the static yield stress measurement; a shear rate of 0.01 s$^{-1}$ is applied to the sample after a rest of 4 min. \label{fig:refY}}
\end{figure}

After the cup is set in the rheometer, the vane tool is slowly inserted, so that the last significant flow undergone by the sample before the start of the rheometry procedure is the one generated when pouring it into the cup. Because of the random nature of this flow, the sample is assumed to be homogeneous and isotropic prior to the start of the measurement sequence. The critical stress to make the samples flow after a 4 min rest (static yield stress $\tau_{ys}$) is first measured by initiating flow from rest: a small and constant shear rate (0.01 s$^{-1}$) is applied to the samples. This small velocity ensures that the contribution of viscous effects to the measured stress is negligible. The stress-strain curve obtained during this measurement on the suspending paste is shown in figure~\ref{fig:refY}. For very small deformation, the stress first increases very abruptly with the deformation, which shows that the critical deformation to exit the linear elastic regime of the metakaolin paste is very small. As the deformation increases, the stress rapidly reaches a peak value $\tau_{ys}$, at which point the material starts to flow. As flow goes on, the stress relaxes to a plateau value. The large stress overshoot at the start of flow is typical of the flow start-up of thixotropic materials, in which a microstructure has built up during the resting time. The slow decrease in stress once flow is established corresponds to the destructuration of the material under flow, which lowers its apparent viscosity~\citep{coussot2005rheometry,  barnes1997thixotropy, ovarlez2008influence}. \\

\begin{figure}[]
\centering
\includegraphics[scale=0.235]{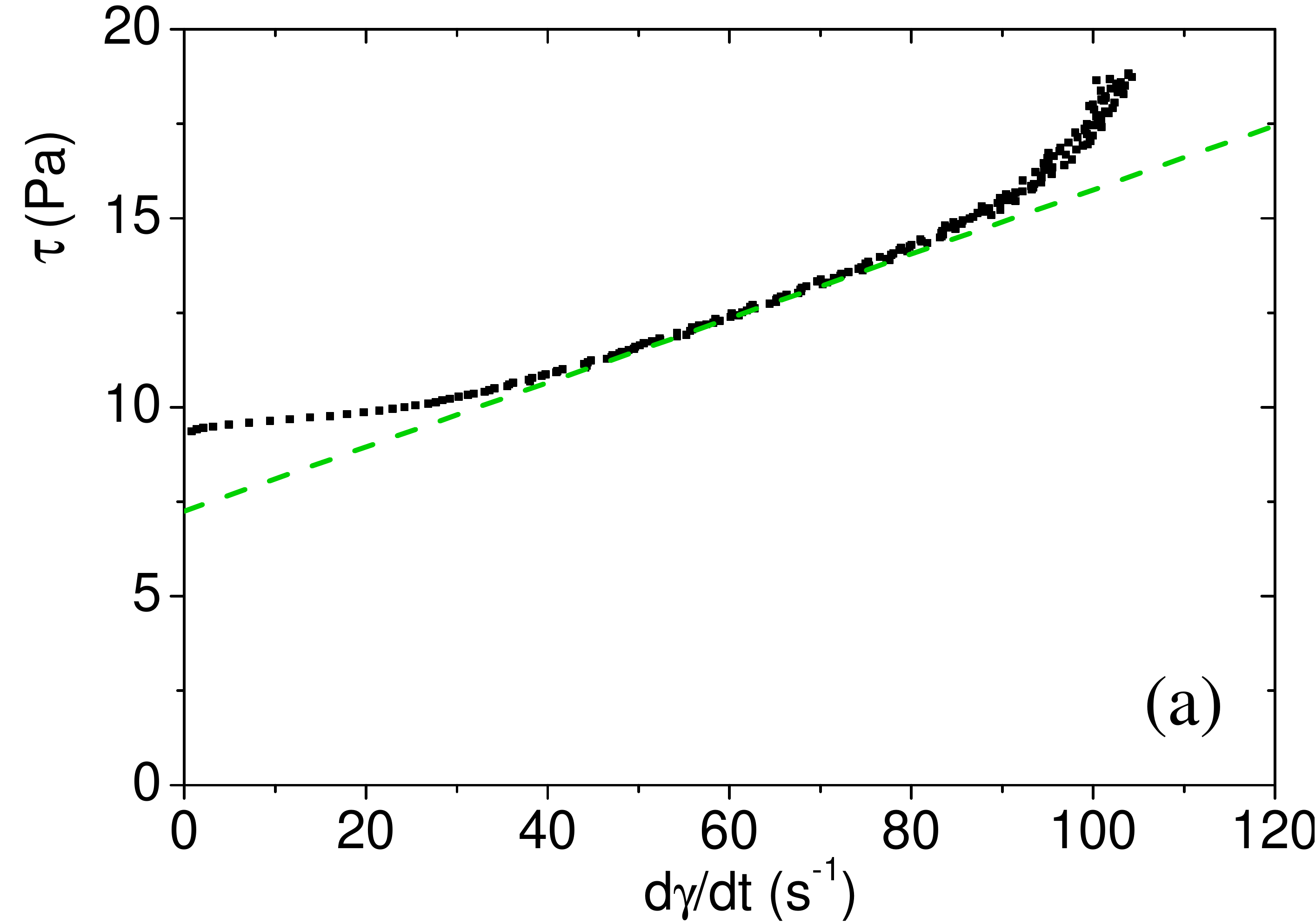}
\includegraphics[scale=0.235]{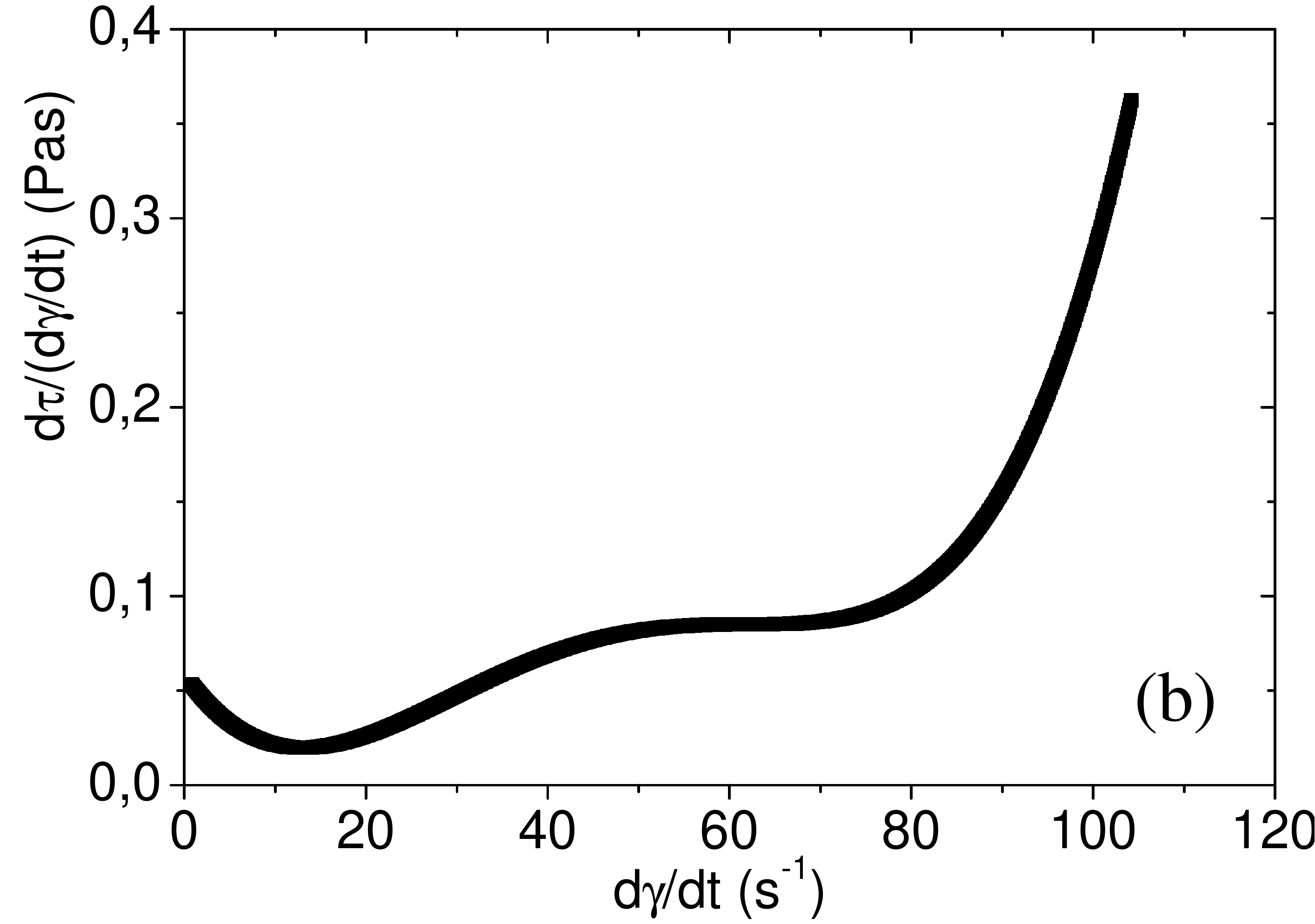}
\caption{Characterization of the flow behavior of the suspending paste: (a) Stress as a function of shear rate obtained during the decreasing shear stress ramp. The dashed line is a linear fit to the data in the region of constant plastic viscosity. (b) Derivative of the flow curve with respect to the flow rate. A viscosity plateau is observed at intermediate shear rates. The onset of shear thickening at higher flow rates is clearly visible. \label{fig:refF}}
\end{figure}

The constant shear rate is applied for 60 s, followed by a rest (zero stress) of 10 s, after which the flow curve of the samples is measured by imposing a shear stress ramp, first increasing from 0 Pa to a maximum stress of the order of twice the static yield stress, and then decreasing back to 0 Pa. For this measurement, it is necessary to work at imposed stress because the metakaolin paste starts flowing very abruptly at yielding, and the feedback loop of the rheometer is then unable to impose the chosen shear rate ramp. The flow curve of the suspending paste during the decreasing shear stress ramp is shown in figure~\ref{fig:refF} (a). At the beginning of the ramp, the stress is high and the paste flows. When stress is decreased, the paste stops flowing at a finite value of applied stress: this value defines the dynamic yield stress $\tau_{yd}$ of the paste. For our metakaolin paste, the dynamic yield stress is much lower than the static yield stress. This can be understood by the fact that it is easier to keep the destructured material flowing than break the structure to initiate flow from rest~\citep{barnes1997thixotropy}. Above the dynamic yield stress, after a smooth transition at low shear rates, the stress seems to increase linearly with the shear rate. At high shear rates, however, the behavior of the paste deviates from the linear regime and a faster increase in the stress seems to indicate that the metakaolin paste is shear thickening. To better evidence this phenomenon, we fit the whole flow curve to a polynomial function and compute its derivative $d\tau/d\dot{\gamma}$, which is the apparent plastic viscosity of the material. The evolution of this apparent plastic viscosity with the shear rate is plotted in figure~\ref{fig:refF} (b). A plateau in the apparent viscosity at intermediate shear rates allows us to define the region for which the behavior of the paste is linear. This plateau is followed by a sharp increase, clarifying that the paste is apparently shear thickening. This apparent shear-thickening is likely due to inertial effects: because the paste's viscosity is rather low, its flow in a Couette-like geometry is prone to develop instabilities. In a Couette cell, the first instability to appear in the flow of low viscosity Newtonian fluids is the Taylor-Couette instability, characterized by the dimensionless Taylor number
\begin{equation}
Ta=\frac{\Omega^2a^3R_m}{\nu^2}
\end{equation}
where $\Omega$ is the rotation rate of the inner cylinder, $a$ is the gap between the two cylinders, $R_m$ is the mean radius of the two cylinders and $\nu$ is the kinematic viscosity of the fluid. $Ta$ is only rigorously defined for a Newtonian fluid flowing in a Couette cell, in which case the instability develops above a critical Taylor number of 1712~\citep{guyon2001physical}. By using this critical Taylor number to get an order of magnitude of the importance of inertial effects in our Bingham fluids flowing in a vane geometry, we can compute a critical shear rate at which the instability is expected. This value is around 40 s$^{-1}$  for the metakaolin paste. We do not expect the agreement to be quantitative, but the order of magnitude for this critical shear rate is compatible with the onset of shear thickening observed in the paste. This confirms our hypothesis that this apparent shear-thickening, also observed in the foamed samples, is of inertial origin and we will thus not consider it when determining the flow behavior of our samples. We fit the linear regime of the flow curve with a Bingham law
\begin{equation}
\tau(\dot{\gamma})=\tau_c+k_1 \dot{\gamma}
\end{equation}
from which we extract the plastic viscosity $k_1$ describing the dissipation in the samples during flow. This law for the suspending paste is plotted with a dashed line in figure~\ref{fig:refF} (a). Because the flow curves of the paste and the foamed samples deviate from the linear behavior at low shear rates, we measure the dynamic yield stress $\tau_{yd}$ directly from the experimental curve, and do not use the value $\tau_c$ provided by the linear fit at higher shear rates.\\

Table~\ref{tab:recap} summarizes the relevant physical quantities to quantify the rheological behavior of the metakaolin suspending paste in our samples, and the bubbles added into it.
\begin{table}
\centering
\begin{tabular}{ccccccc}
\hline
& $\tau_{ys}(0)$ & $\tau_{yd}(0)$ & $k_1$ & $\sigma$ & $R_b$ & $2\sigma/R_b$ \\
& (Pa) & (Pa) & (Pas) & (mN/m) & ($\mu$m) & (Pa) \\
\hline
series 1 & 25 $\pm$ 2 & 10.2 $\pm$ 0.5 & 0.082 $\pm$ 0.002 & 35 $\pm$ 1 & 35 $\pm$ 10 & 2000 $\pm$ 500 \\
series 2 & 25 $\pm$ 2 & 10.2 $\pm$ 0.5 & 0.082 $\pm$ 0.002 & 35 $\pm$ 1 & 100 $\pm$ 20 & 700 $\pm$ 100 \\
\hline
\end{tabular}
\caption{Properties of the suspending metakaolin paste and the bubbles added into it. $\sigma$ is the surface tension of the aqueous phase.\label{tab:recap}}
\end{table}

\section{Experimental results}
\subsection{Critical stress to induce flow}
\begin{figure}[]
\centering
\includegraphics[scale=0.23]{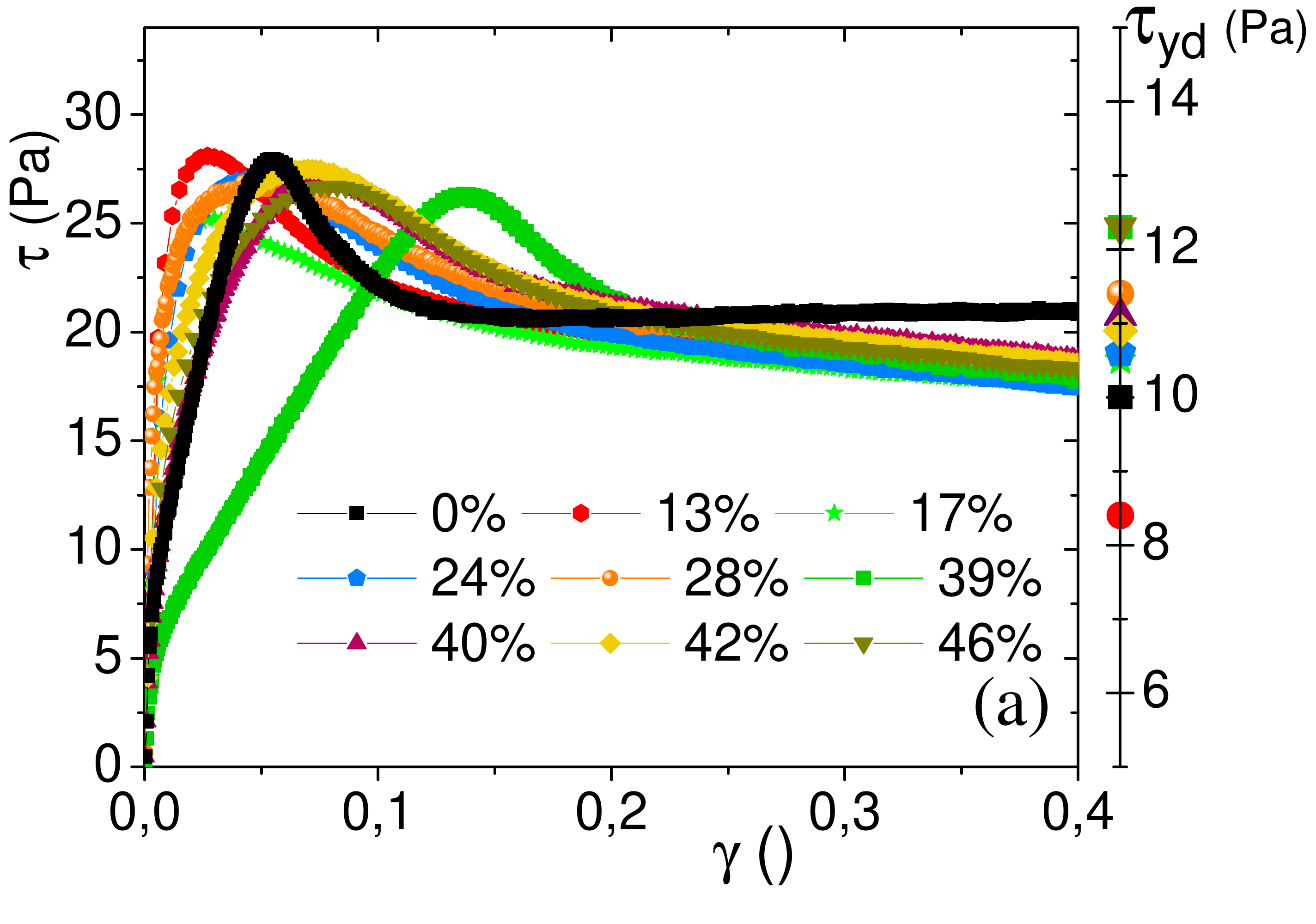}
\includegraphics[scale=0.23]{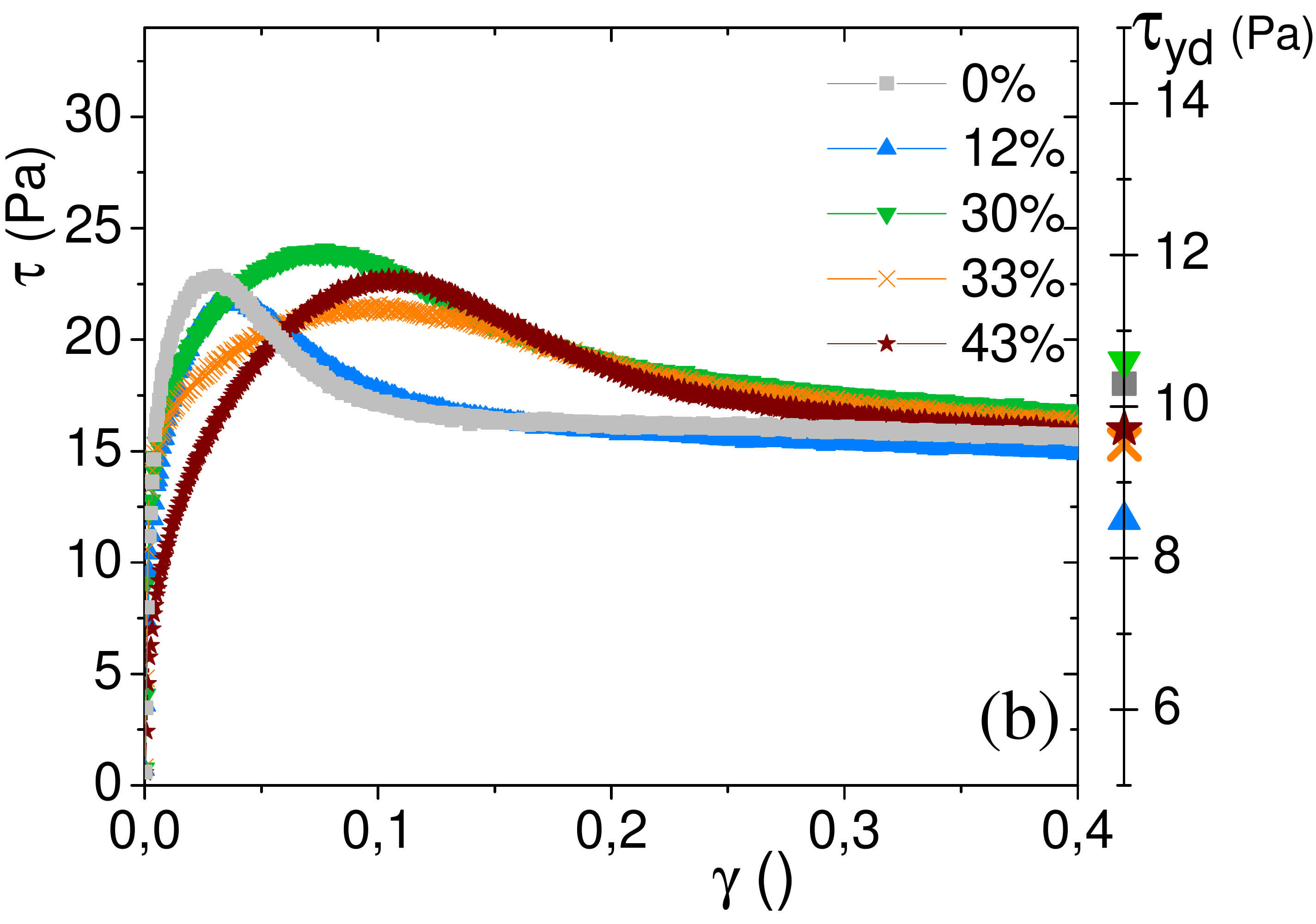}
\caption{Stress as a function of strain during the static yield stress measurement on suspensions of (a):   $R_b=35 \mu$m and (b): $R_b=100 \mu$m bubbles in the metakaolin paste. A shear rate of 0.01 s$^{-1}$ is applied to the samples after a 4 minute rest. The dynamic yield stress of the same samples, obtained from the flow curves, is plotted on the right axis. \label{fig:yield}}
\end{figure}

\begin{figure}[]
\centering
\includegraphics[scale=0.25]{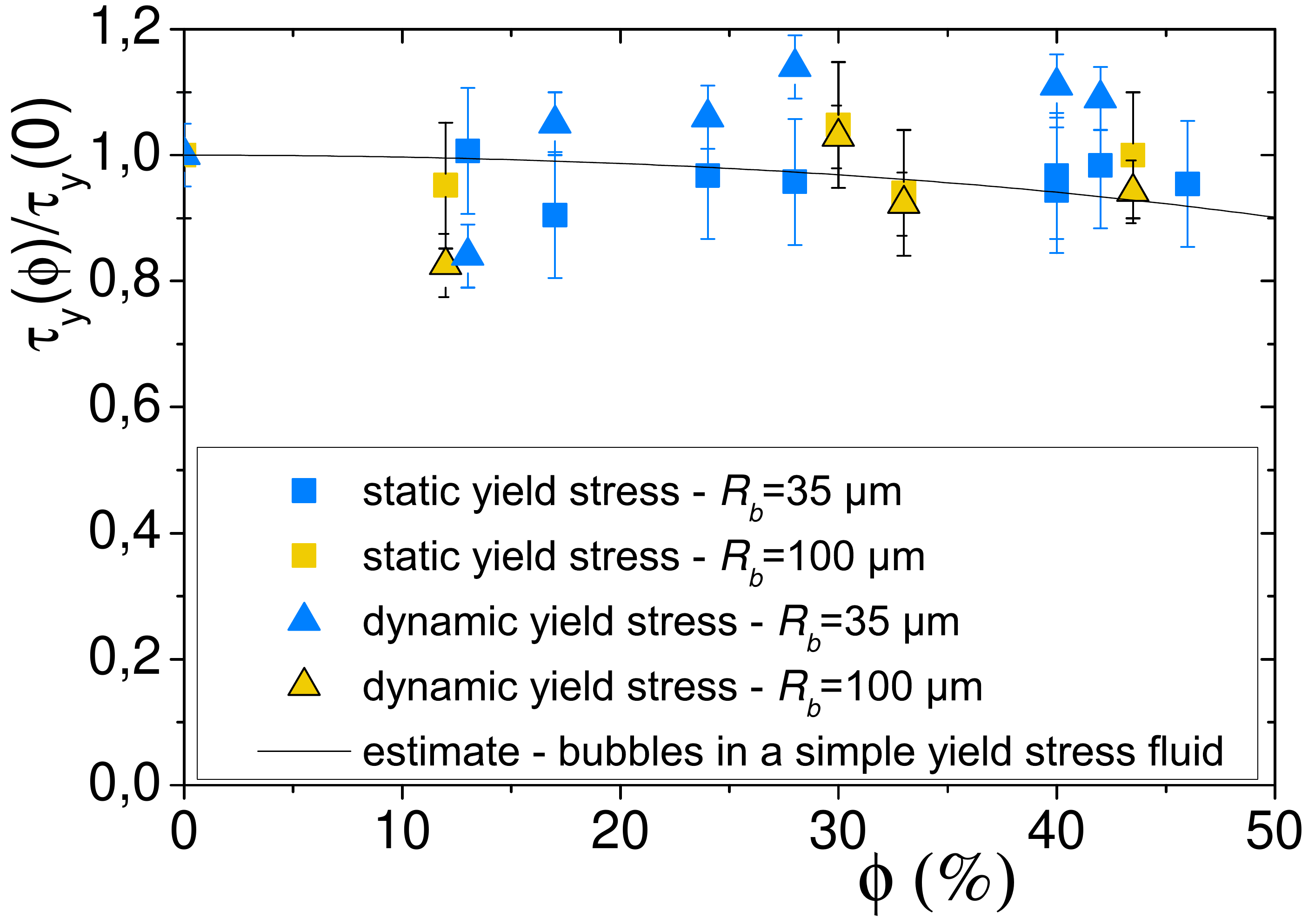}
\caption{Dimensionless yield stress (static: $\Box$, and dynamic: $\triangle$) as a function of the gas volume fraction, for the two series of suspensions studied. The full line is a micro-mechanical estimate for the dimensionless static yield stress of suspensions of monodisperse bubbles in model yield stress fluids~\citep{ducloue2015rheological}. \label{fig:yieldnondim}}
\end{figure}

The suspending metakaolin paste is, as we have shown along the rheometry procedure, a yield stress fluid: a minimum stress $\tau_{ys}$ is required to make it flow. We first focus on flow start-up of the foamed samples. The stress-strain curves obtained by imposing a small and constant velocity to the samples after a rest of 4 minutes are presented for the two series of foamed samples studied in figure~\ref{fig:yield}.
We observe that the foamed samples exhibit the same behavior as the suspending metakaolin paste, with a sharp increase in stress at small deformation, followed by a stress overshoot and a subsequent relaxation. The peak value of stress is the same for all the samples within experimental reproducibility, and is equal to the one of the suspending paste. The introduction of bubbles does not seem to affect the static yield stress of the samples. \\
The dynamic yield stress of the two series of samples, obtained from the flow curves, is plotted on the right axis of each subfigure in figure~\ref{fig:yield}. It is much lower than the static yield stress of the samples, and comparable to $\tau_{yd}$ of the suspending paste, with a scatter but no trend in its dependence on the gas volume fraction $\phi$. The dynamic yield stress, too, seems to be insensitive to the presence of the bubbles within the range of experimental parameters investigated here.\\
To better explicit the dependence of the yield stress, static $\tau_{ys}(\phi)$ and dynamic $\tau_{yd}(\phi)$, on $\phi$, we now analyze the data by plotting the static yield stress (respectively the dynamic yield stress) of the samples, scaled by $\tau_{ys}(0)$ (resp. $\tau_{yd}(0)$), as a function of $\phi$. The result of this operation is shown for the two series of foamed samples by the square symbols (resp. the triangular symbols) in figure~\ref{fig:yieldnondim}.

As expected from the experimental curves obtained during the measurements, both the dimensionless static yield stress and the dimensionless dynamic yield stress are approximately constant, equal to 1. In the range of bubble sizes and void volume fraction studied, the static and dynamic yield stress of the foamed samples are imposed by the suspending metakaolin paste.\\

\subsection{Dissipation during flow}
\begin{figure}[]
\centering
\includegraphics[scale=0.235]{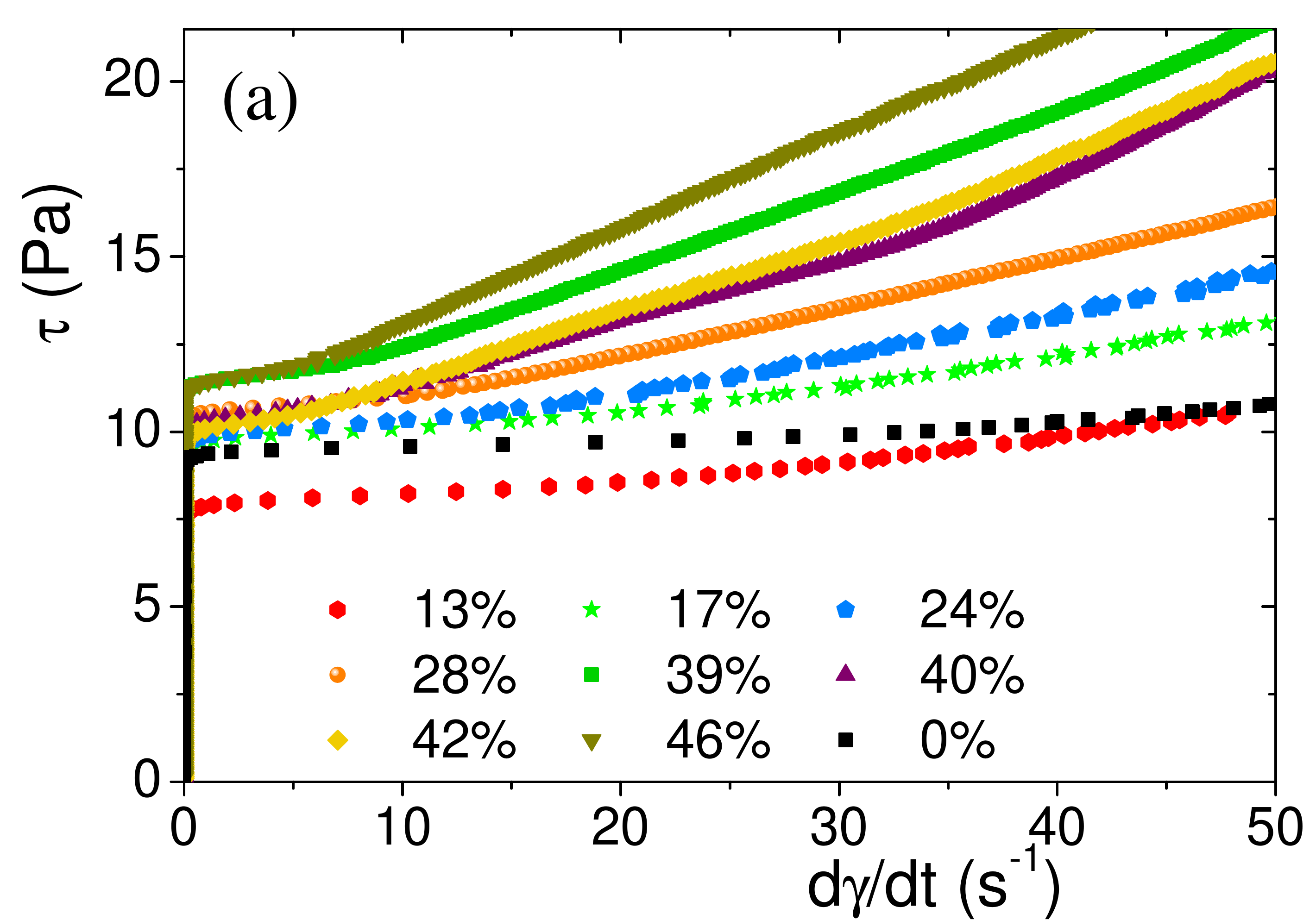}
\includegraphics[scale=0.235]{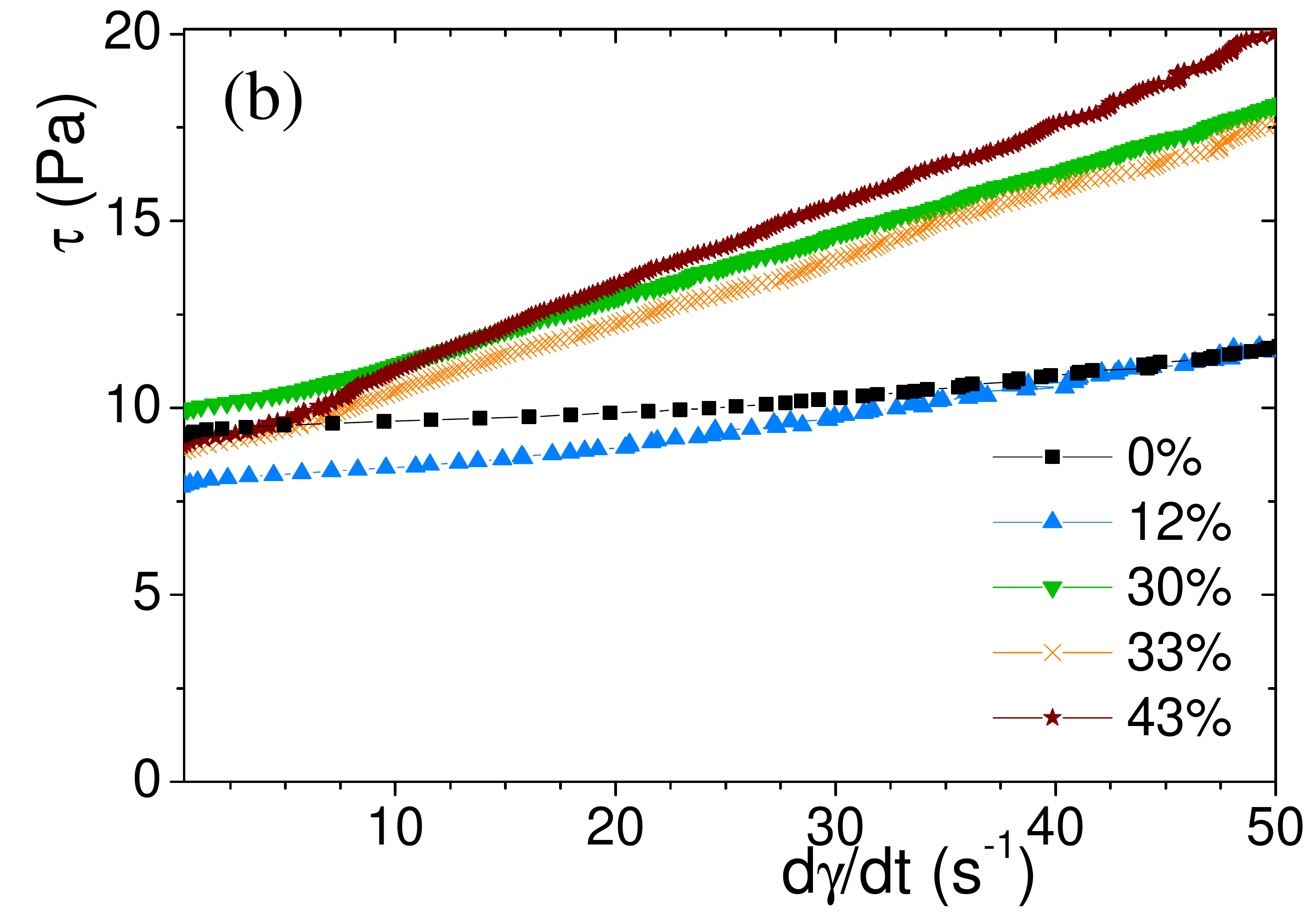}
\caption{Flow curves of the two series of suspensions studied: (a) $R_b=35 \mu$m; (b) $R_b=100 \mu$m. \label{fig:flow}}
\end{figure}

\begin{figure}[]
\centering
\includegraphics[scale=0.25]{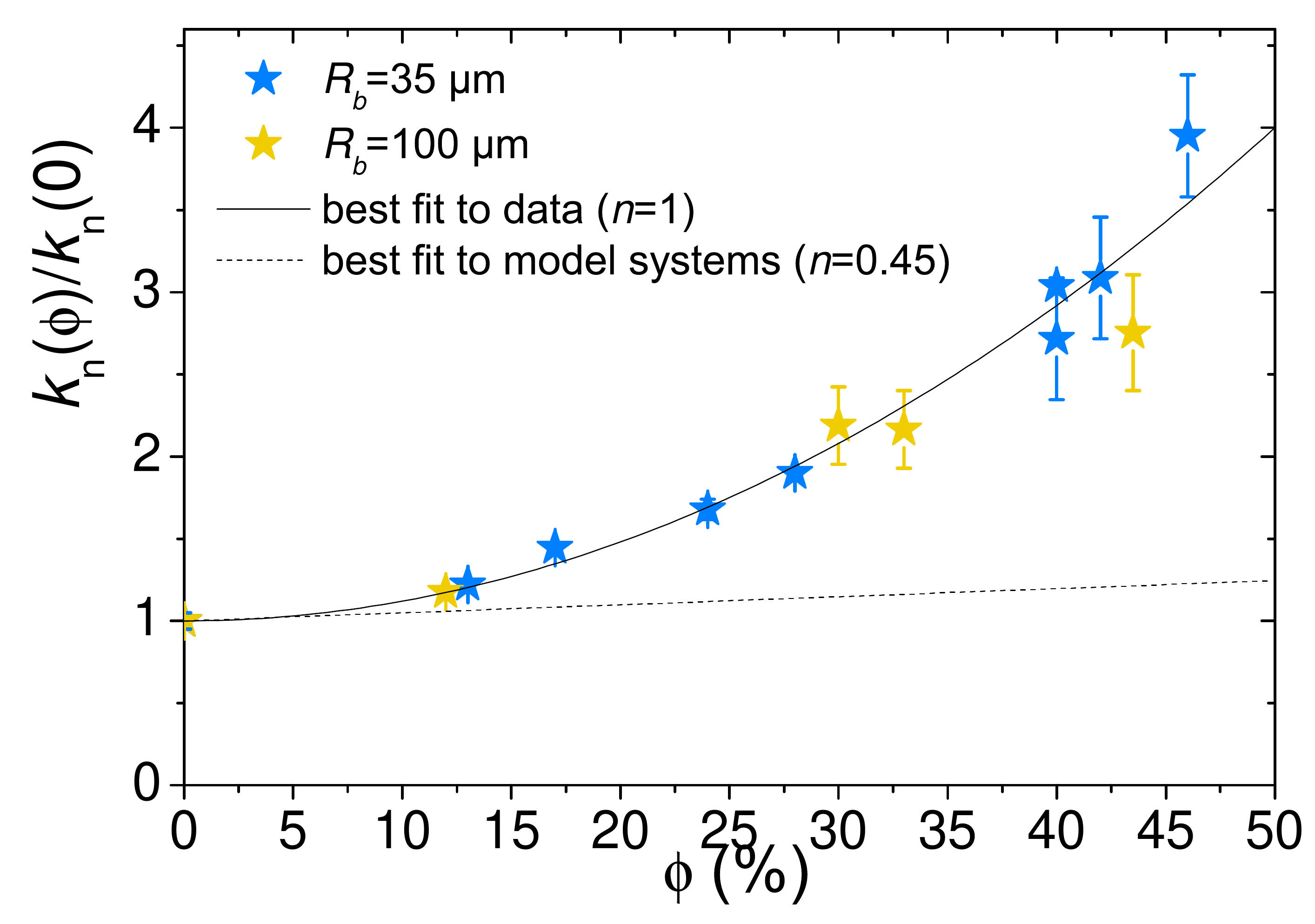}
\caption{Dimensionless plastic viscosity of the two series of suspensions  of bubbles in the metakaolin paste ($n=1$), as a function of the gas volume fraction.  The full line is a phenomenological best fit to the data. The dashed line is a best fit to data obtained on suspensions of bubbles in model yield stress fluids with $n=0.45$.\label{fig:flownondim}}
\end{figure}

For stresses higher than the yield stress, the suspending metakaolin paste flows, with a relationship between the shear rate and shear stress which is non-linear and has been described in section~\ref{section:rheo}. The flow curve obtained during the decreasing ramp of shear stress on the suspending metakaolin paste and the foamed samples is shown for the two series of suspensions in figure~\ref{fig:flow}. For both series of suspensions, the general shape of the curve is not modified by the presence of the bubbles: there is a finite dynamic yield stress, followed by a linear regime at higher shear rates which can be fitted to a Bingham model. We observe that the plastic viscosity $k_1$ increases with the gas volume fraction. \\
The presence of the air bubbles, although of little or no effect on the dynamic yield stress, clearly increases the plastic viscosity of the foamed samples during flow. To quantify this increase in the flow dissipation, we now plot in figure~\ref{fig:flownondim} the plastic viscosity $k_1(\phi)$ of the foamed samples scaled by the one of the suspending metakaolin paste $k_1(0)$. As expected from the flow curves of the samples, the dimensionless plastic viscosity strongly increases with $\phi$: 40\% of bubbles in the paste is enough to triple the plastic viscosity. 

\section{Discussion}
\subsection{Yield stress}
The somewhat surprising result that the introduction of 50\% of bubbles in the paste does not change its yield stress can be understood in the light of work done on model foamed materials. The behavior of suspensions of bubbles in model yield stress fluids has been experimentally studied, in the same range of void volume fraction as here, by using suspensions of bubbles in concentrated oil-in-water emulsions~\citep{ducloue2015rheological}. Those emulsions are non-reactive simple yield stress fluids~\citep{mason1996yielding, ovarlez2008wide}: their rheological behavior does not depend on the flow history. This well-controlled behavior justifies their use as model yield stress fluids. It has been shown that the yield stress of these foamed emulsions is equal to the one of the suspending emulsion providing that capillary effects are strong enough in the system. More precisely, the rheology of the foamed simple yield stress fluid is the result of the interplay of the two kinds of stresses acting on the bubbles when the sample starts to flow. On the one hand, the just yielded suspending fluid exerts a stress on the bubbles, which is of the order of magnitude of the fluid's yield stress and tends to deform the bubbles. On the other hand, capillarity acting on the surface of the bubbles tends to minimize to bubbles' surface area by restoring a spherical shape. The competition of these two mechanisms determines the relative rigidity of the bubbles in the just yielded fluid and can be quantified by introducing a plastic capillary number:
\begin{equation}
Ca_{\tau}=\frac{\tau_y(0)}{2\sigma/R_b}
\end{equation}
which is the ratio of the paste yield stress $\tau_y(0)$ to the capillary stress $2\sigma/R_b$. When surface tension forces largely dominate over the yield stress of the suspending paste, this capillary number is very low and the bubbles behave as rigid spherical objects in the paste: the stress in the paste is not high enough to deform them. Note however that they differ from rigid particles in the boundary condition with the fluid, which slips on the bubbles' surface. As the relative importance of surface tension is decreased, $Ca_{\tau}$ increases and the bubbles become increasingly soft in the paste. \\

Experiments have shown that below a capillary number of approximately 0.1, the yield stress of the samples is not affected by the presence of the bubbles. For the model systems of monodisperse bubbles in simple yield stress fluids at moderate void volume fraction (well below the percolation threshold of 64\%), it is possible to quantitatively describe and estimate the behavior of the foamed samples at zero capillary number thanks to homogenization techniques~\citep{doi:10.1061/9780784412992.224}. The result of the estimation is in good agreement with the experimental data obtained on those model systems and is plotted with a full line in figure~\ref{fig:yieldnondim}.\\
In the case of the foamed metakaolin samples, the plastic capillary numbers are 0.01 and 0.03 for the $R_b=35$ $\mu$m and $R_b=100$ $\mu$m samples, respectively. From the results on foamed model yield stress fluids, it is then expected that the yield stress of the foamed metakaolin samples should not depend on $\phi$ in the range of volume fractions investigated. The constant yield stress of the foamed metakaolin samples shows that their behavior at yielding can be predicted in the same way as in the model systems, although the rheology of the suspending paste is in fact more complicated. \\
Experiments on suspensions of bubbles in model yield stress fluids have also shown that for higher plastic capillary numbers, the bubbles become deformable at yielding of the suspending paste. At the scale of the samples, the softness of the bubbles in the fluid results in a drop of the yield stress with $\phi$. For large deformations, it has also been observed experimentally that around the same critical capillary number (approximately 0.5 for the suspensions in the concentrated emulsions), bubbles rupture during mixing with the paste~\citep{kogan2013mixtures}. We did not investigate this regime for metakaolin pastes but the same features can be expected as in model materials. The plastic capillary number is thus also of particular interest in any application requiring the preservation of bubble size during the flow of a foamed metakaolin paste: it gives an estimate of the maximum shear stress that can be applied to the system without breaking the bubbles.\\

\subsection{Plastic viscosity}
The plastic viscosity of the foamed samples dramatically increases with $\phi$ for both series of suspensions studied. Interestingly, data for both series of suspensions fall on the same curve, which suggests that the bubble size is not a relevant parameter in this viscosity increase. Indeed, as we have explained in the experimental procedure, the maximum stress applied during the shear stress ramp is about twice the static yield stress. We have shown in the discussion of the static yield stress measurements that $\tau_{ys}(0)$ is much lower than the capillary stress scale, resulting in the bubbles being non-deformable in the suspending metakaolin paste. During the flow curve measurement, the scale of the capillary stress will as a consequence also be very large compared to the stress scale in the metakoalin paste, meaning that the bubbles remain rigid in the flow within the range of shear rates considered here. \\

We can then quantify the relative increase in the plastic viscosity for suspensions of rigid bubbles in the metakaolin paste by fitting our experimental data with a polynomial function. The result of this fit is plotted with a full line in figure~\ref{fig:flownondim}. The increase for the metakaolin samples is quadratic (equation of the curve: $k_1(\phi)/k_1(0)=1+12\phi^2$). For comparison, the relative increase in the plastic viscosity of suspensions of rigid bubbles in model yield stress fluids is plotted with a dotted line in figure~\ref{fig:flownondim} (best fit to the experimental data obtained on those systems). The increase is much less, but the results on the metakaolin paste and the concentrated emulsion cannot be compared directly because the flow behavior of the emulsion has a different power law than the metakaolin paste: its flow curve is well fitted to a Herschel-Bulkley model $\tau(\dot{\gamma})=\tau+k \dot{\gamma}^{n}$ with $n=0.45$. For this system, the measured increase in plastic viscosity is linear: $k_{0.45}(\phi)/k_{0.45}(0)=1+0.49\phi$. The difference in the rheology of the suspending fluid is significant, and can account for the two different responses for the plastic viscosity with the volume fraction. Indeed, the global dissipation in the foamed samples is very sensitive to the way shear is localized between the bubbles during flow, and this local behavior of the suspending fluid is determined by the flow curve of the suspending fluid. \\

\section{Conclusions}
We have prepared model foamed metakaolin samples, in which the rheology of the suspending metakaolin paste, the bubble size and the bubble volume fraction are controlled. Two series of samples have been studied, with two different mean bubble radii. We have shown thanks to careful rheology measurements that the static and dynamic yield stress of the foamed samples are the same as the ones of the suspending paste for the bubble sizes we have investigated. This result can be understood in terms of bubble deformability, quantified by a capillary number. Our rheology experiments have also shown that the presence of the bubbles in the paste increases the dissipation during flow. A comparison with suspensions of bubbles in model yield stress fluids shows that it is also the case in those systems, and that it is likely that dissipation increases faster for suspensions in fluids of larger plastic index. We have given phenomenological quantification of this dissipation in the foamed model fluid and the foamed metakaolin samples.\\
Those results are very promising for the understanding of the rheology of foamed metakaolin pastes, but more importantly have proven that the framework developed for model foamed yield stress fluids can also apply to more complex systems. This generality could make it relevant to the flow and processing of other foamed mineral pastes, such as foamed clay, concrete or plaster slurries. \\

\section*{Ackowledgments}
We are thankful to AGS Min\'eraux (Imerys group) for kindly providing us with the metakaolin as a sample. Financial support from Saint-Gobain Recherche (Aubervilliers, France) is gratefully acknowledged. 

\bibliographystyle{elsarticle-harv} 
\bibliography{biblioK}

\end{document}